# UNDERSTANDING THE DYNAMICS OF INFORMATION FLOW DURING DISASTER RESPONSE USING ABSORBING MARKOV CHAINS

Yitong Li
Wenying Ji

Department of Civil, Environmental, and Infrastructure Engineering
George Mason University
Fairfax, VA 22030, USA

## ABSTRACT

This paper aims to derive a quantitative model to evaluate the impact of information flow on the effectiveness of disaster response. At the core of the model is a specialized absorbing Markov chain that models the process of delivering federal assistance to the community while considering stakeholder interactions and information flow uncertainty. Using the proposed model, the probability of community satisfaction is computed to reflect the effectiveness of disaster response. A hypothetical example is provided to demonstrate the applicability and interpretability of the derived quantitative model. Practically, the research provides governmental stakeholders interpretable insights for evaluating the impact of information flow on their disaster response effectiveness so that critical stakeholders can be targeted proactive actions for enhanced disaster response.

## 1  INTRODUCTION

Often resulting in massive devastation, natural disasters are major factors affecting community resilience. From 2008-2018, natural disasters caused $850 billion of economic losses in the U.S. and $1.5 trillion around the world (NSF 2019). To ensure the community remain functional, timely and appropriate responses are required for addressing the community needs (e.g., highway drainage during flooding (Chen et al. 2020) and shelter construction during hurricanes (Sanusi et al. 2020). During disaster response, the supporting role (e.g., technical, financial, or professional supports) of federal government becomes extremely vital in situations when major disasters exceed the response capacity of the state and local government (FEMA 2019). To ensure an efficient delivery of federal assistance to the community, governmental stakeholders at various levels (i.e., state, federal, and local) need to work together as a collaborative network (Nowell et al. 2017; FEMA 2019). Previous research has identified three main attributes contributing to an effective stakeholder collaborative network—the implementability of disaster response guidelines, the capability of the network to adapt to changing response needs, and the effective information flow across stakeholders (Nowell et al. 2017)—among which information flow is the most impactful one since effective responding actions always require timely information (Kapucu and Garayev 2013; O'Leary and Vij 2012).

In the domain of disaster management, previous research on information flow has focused on improving the quality of information flow through information communication technology (ICT). Examples include the design of an integrated communication and information system for disaster response and recovery (Meissner et al. 2002) and the analysis of technical factors on the application of communication technology in emergency planning (Dilmaghani and Rao 2007). Although enhancing information flow from the technological perspective is fundamental, understanding the patterns of information flow among stakeholders is also critical (Sagun et al. 2009; Kapucu and Garayev 2013). To achieve this, network analyses have been widely applied to explore factors contributing to effective interorganizational communication (Kapucu 2006) and to model stakeholder communication networks using social media data (Rajput et al. 2020). Although capable of obtaining insights for improving disaster response, these studies mainly focused on reflecting stakeholder interactions of information sharing and ignored the details of information flow (e.g., information flow frequency), which impedes the discovery of in-depth knowledge on how information flow impacts disaster response. In addition, during disaster response, the continuously changing needs further add uncertainties to the information flow. For example, to address a new response requirement, information may flow between stakeholders who never had interactions before. Therefore, to enhance the understanding of how information flow impacts disaster response, a quantitative model that is capable of modeling complex stakeholder interactions and incorporating information flow uncertainty is needed.

The objective of this research is to derive a quantitative model to evaluate the impact of information flow on the effectiveness of disaster response. At the core of the quantitative model is a specialized absorbing Markov chain that models the responding process while incorporating stakeholder interactions and information flow uncertainty. Through the specialized model, the probability of community satisfaction is derived to measure the effectiveness of disaster response. The remainder of this paper is organized as follows. In the next section, a simplified conceptual model is demonstrated to illustrate the responding process and the types of information flow involved. Built upon the conceptual model, a specialized absorbing Markov model is presented, and a step-by-step explanation is given in the methodology section. Following the methodology section, a hypothetical example is provided to demonstrate the applicability and interpretability of the developed model. In the end, contributions, limitations, and future work are concluded.

## 2    CONCEPTUAL MODEL

In this section, a simplified conceptual model is illustrated (see Figure 1) to describe the responding process and involved types of information flow. The model is constructed as per the disaster response process described in the National Response Framework (FEMA 2019) and governmental stakeholder relationships (Lindell et al. 2006). Essentially, there are three levels of stakeholders involved in disaster response, i.e., federal, state, and local levels. Here, stakeholders are defined as governmental agencies (e.g., Federal Emergency Management Agency and Department of Homeland Security are federal stakeholders) that are responsible for disaster response.

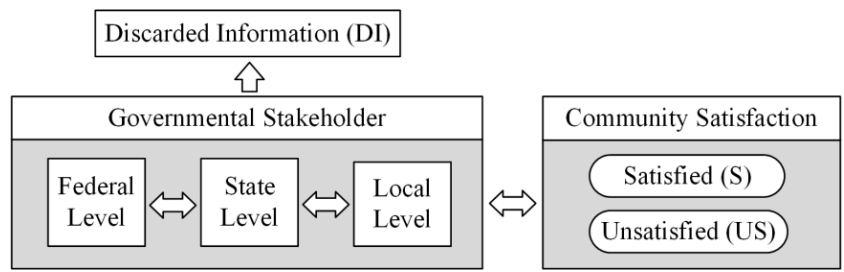

Figure 1: Process of delivering federal assistance to the community (arrows indicate directions of information flow).

During the process of delivering federal assistance to the community, federal stakeholders provide assistance (e.g., federal loan program) for addressing community needs, state and local stakeholders support disaster response based on their capabilities (e.g., roles, responsibilities, and resource availabilities). In this process, all exchanged information among stakeholders for the purpose of disaster response is considered as information flow. In response to these actions, the community reflects satisfaction if federal assistance is effectively delivered to address the community needs, and vice versa. The community satisfaction is obtainable through measuring the public sentiment through social media platforms (Chen et al. 2019).

Based on the process, two categories of information flow are classified: bidirectional and unidirectional. In detail, bidirectional information flow represents situations when collaborative efforts are needed to achieve an efficient disaster response. This type of information flow exists between stakeholders either at different levels or same levels. Unidirectional information flow represents situations when state and federal stakeholders discard the received information due to failures of taking actions (e.g., failure of interpretation and incapability of taking actions). This type of information flow is considered as ineffective information flow, and it exists between governmental stakeholders and the discarded information state.

In summary, the process of delivering federal assistance to the community starts from federal stakeholders and ends in any of the following scenarios: 1) information related to federal assistance is discarded (DI) during the process, 2) the community receives federal assistance and is satisfied (S), and 3) the community receives federal assistance but is unsatisfied (US). The conceptual model defined in this section will be used to derive the quantitative model.

## 3    MODEL DEVELOPMENT

In this section, a specialized absorbing Markov chain model, which incorporates stakeholder interactions and information flow uncertainty, is derived to model the process of delivering federal assistance to the community. Using the model, the probability of community satisfaction ($P_S$) is computed to measure the effectiveness of disaster response (i.e., the level of community satisfaction with federal assistance).

An absorbing Markov chain describes a stochastic process that begins from one transient state and moves, successively, from its current transient state $i$ to another state $j$ with probability $\theta_{ij}$. The process stops once it ends up in an absorbing state (i.e., $\theta_{ii} = 1$) (Resnick 2013). During the process of delivering federal assistance to the community, the information flow starts from the federal stakeholder, then passes through several state and local stakeholders, and eventually ends at the community or the discarded information state. The information flow process aligns with the mathematical property of the absorbing Markov chain; therefore, it is selected to model the responding process. The specialized absorbing Markov-chain model is illustrated in Figure 2.

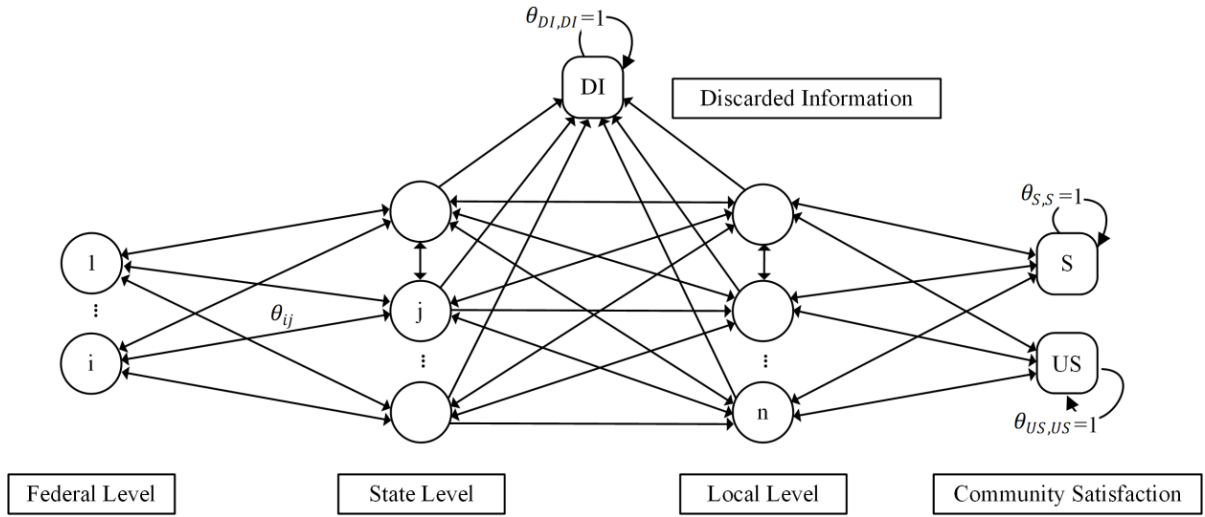

Figure 2: Specialized absorbing Markov chain model.

Once a new type of federal assistance is declared, it is assumed that $n$ stakeholders are involved in delivering assistance to the community. These $n$ stakeholders consist $n$ transient states (circle nodes) of the specialized absorbing Markov chain. The three absorbing states (square nodes) DI, S, and US represent end scenarios of disaster response. A transient state $i$ indicates that stakeholder $i$ has received information, and after taking actions, stakeholder $i$ flows information to stakeholder (or any absorbing states) $j$ with probability $\theta_{ij}$. In this research, $\theta_{ij}$ indicates the extent to which stakeholder $i$ flow information to stakeholder $j$ (or any absorbing states). The higher the probability is, the more likely stakeholder $i$ flows information to stakeholder $j$ (or any absorbing states). Assuming stakeholder $i$ has $K$ interacting states (i.e., stakeholders and absorbing states). Frequencies of information flow from stakeholder $i$ to each of its interacting states are obtainable from historical disaster response. Using the information flow frequency, the probabilities of information flow from stakeholder $i$ to each of its interacting states are computed using the multinomial distribution. The multinomial distribution is used to infer the probabilities that an event with $K$ outcomes comes up with a certain outcome. Let $\mathcal{D} = \{N_{i1}, ..., N_{iK}\}$ be a random vector, where $N_{ij}$ is the frequency of information flow from stakeholder $i$ to its interacting state $j$ and $n_i$ is the total frequency of information flow from stakeholder $i$ to all its interacting states, the probabilities of information flow from stakeholder $i$ to any of its interacting states ($\boldsymbol{\theta} = (\theta_{i1}, \theta_{i2}, ..., \theta_{iK})$) are modeled as a multinomial distribution with the probability mass function expressed in equation (1) (Murphy 2012).

$$p(\boldsymbol{\theta}|\mathcal{D}) = \binom{n_i}{N_{i1}, ..., N_{iK}} \prod_{j=1}^{K} \theta_{ij}^{N_{ij}} \qquad (1)$$

Where $\sum_{j=1}^{K} \theta_{ij} = 1$ and $\sum_{j=1}^{K} N_{ij} = n_i$.

During disaster response, due to the uncertain characteristics of natural disasters (Kapucu and Garayev 2013; Kapucun et al. 2010), information flow among stakeholders vary. To appropriately incorporate information flow uncertainty, the Bayesian statistics-based Dirichlet-multinomial model is used. Bayesian statistics is a systematic way of updating parameters of interest (i.e., posterior distribution) by combining both previous knowledge (i.e., prior distribution) and newly observed data (i.e., likelihood distribution) (Gelman 2013). In the model, the prior distribution is the Dirichlet distribution that represents the prior knowledge of the probabilities of information flow from stakeholder $i$ to any of its interacting states. The likelihood distribution is the multinomial distribution that represents the newly observed frequencies of information flow from stakeholder $i$ to any of its interacting states. Since the Dirichlet distribution is a conjugate prior for the multinomial distribution, the posterior distribution is also a Dirichlet distribution and is expressed as:

$$p(\boldsymbol{\theta}|\mathcal{D}) = \text{Dir}(\boldsymbol{\theta}|\alpha_1 + N_{i1}, ..., \alpha_K + N_{iK}) \qquad (2)$$

Where $\boldsymbol{\alpha} = (\alpha_1, ..., \alpha_K)$ are shape parameters that control the prior distribution. To remove the effect of external information on current data, noninformative prior distribution $\text{Dir}(\mathbf{1})$ is used (Berger 2013). However, in reality, the prior distribution is determined by the expert's belief, historical data, and existing knowledge. Using $\text{Dir}(\mathbf{1})$, the posterior distribution of the probabilities of information flow from stakeholder $i$ to any of its interacting states is now expressed as:

$$p(\boldsymbol{\theta}|\mathcal{D}) = \text{Dir}(\boldsymbol{\theta}|1 + N_{i1}, ..., 1 + N_{iK}) \qquad (3)$$

Using newly observed frequencies of information flow during disaster responses, the probabilities of information flow are dynamically updated to obtain more accurate and reliable estimations of the probabilities of information flow. Using the calculated probabilities of information flow, the transition matrix $\mathbf{P}$ (in canonical form) of the specialized absorbing Markov chain model, which records the probabilities of information flow between any pair of states, is expressed in equation (4).

$$\mathbf{P} = \begin{pmatrix} \theta_{11} & \theta_{12} & \cdots & \theta_{1n} & \theta_{1DI} & \theta_{1S} & \theta_{1US} \\ \theta_{21} & \theta_{22} & \cdots & \theta_{2n} & \theta_{2DI} & \theta_{2S} & \theta_{2US} \\ \vdots & \vdots & \ddots & \vdots & \vdots & \vdots & \vdots \\ \theta_{n1} & \theta_{n2} & \cdots & \theta_{nn} & \theta_{nDI} & \theta_{nS} & \theta_{nUS} \\ \hline 0 & 0 & \cdots & 0 & 1 & 0 & 0 \\ 0 & 0 & \cdots & 0 & 0 & 1 & 0 \\ 0 & 0 & \cdots & 0 & 0 & 0 & 1 \end{pmatrix}$$

$\mathbf{Q}$   $\mathbf{R}$

$\mathbf{O}$   $\mathbf{I}$

(4)

In the transition matrix, $\mathbf{Q}$ is an $n$-by-$n$ matrix that contains probabilities of information flow between any pair of stakeholders, $\mathbf{R}$ is a $n$-by-3 matrix that contains probabilities of information flow between stakeholders and any absorbing states. $\mathbf{O}$ is a 3-by-$n$ zero matrix and $\mathbf{I}$ is a 3-by-3 identity matrix.

Based on the specialized transition matrix $\mathbf{P}$, the probabilities that disaster response ends up in any absorbing states is computed as follows (Resnick 2013).

$$\mathbf{B} = (\mathbf{I} - \mathbf{Q})^{-1}\mathbf{R} = \begin{pmatrix} P_{1DI} & P_{1S} & P_{1US} \\ P_{2DI} & P_{2S} & P_{2US} \\ \vdots & \vdots & \vdots \\ P_{nDI} & P_{nS} & P_{nUS} \end{pmatrix}$$

(5)

Here, $\mathbf{B}$ is an $n$-by-3 matrix in which each entry represents the probability that disaster response ends up in any absorbing states if it starts from stakeholder $n$. Specifically, the first column records the probabilities of ending up in absorbing state DI. The second and the third columns record the probabilities of ending in absorbing states S and US, respectively. The probabilities of disaster response ending up in any absorbing states are related to each other and the summation of the three probabilities equals 1. In this research, the probabilities of ending up in the absorbing state S ($P_S$), which reflects the level of community satisfaction, is used as a metric to evaluate the effectiveness of disaster response. $P_S$ ranges from 0 to 1, and a larger value indicates more effective disaster response.

## 4   HYPOTHETICAL EXAMPLE

In this section, a hypothetical example is provided to demonstrate the applicability of the proposed quantitative model. The hypothetical example describes a responding process and incorporates all fundamental components (i.e., stakeholders at all levels, information flow frequency, direction of information flow, community satisfaction, and discarded information state) to demonstrate the model's applicability, and to verify the model's performance. The reason of using a simplified hypothetical example is to better explain and demonstrate how the model works. In the example, federal stakeholder A provides a new type of assistance; state stakeholders B and C, and local stakeholders D and E are involved in disaster response. To ensure the successful delivery of federal assistance to the community, a total number of 100 pieces of information are generated from federal stakeholder A. For simplification purposes, information flow demonstrated in the example is unidirectional. The detailed stakeholder network and frequencies of information flow between stakeholders are illustrated in Figure 3.

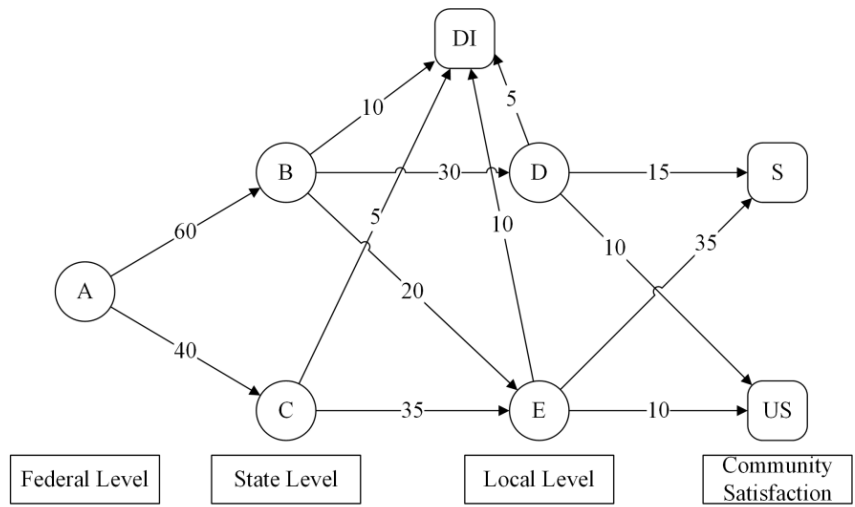

Figure 3: Involved stakeholders and frequencies of information flow.

Given the frequency of information flow, equation (3) is used to compute the probabilities of information flow from one stakeholder to its interacting states. For example, stakeholder B has three interacting states: stakeholder D, stakeholder E, and absorbing state DI. The frequencies of information flow are 30 from stakeholder B to D and 20 from stakeholder B to E. The frequency of ineffective information flow of stakeholder B (i.e., from stakeholder B to absorbing state DI) is 10. Using equation (3), the distribution, which represents the probabilities of information flow from stakeholder B to its interacting states, has the form $p(\theta_{BD}, \theta_{BE}, \theta_{BDI}|\mathcal{D}) = \text{Dir}(31, 21, 11)$. An illustration of $\text{Dir}(31, 21, 11)$ is shown in Figure 4, in which the dark color represents low probability density, while the light color represents high probability density. One random sample selected from $\text{Dir}(31, 21, 11)$ has the form (0.412, 0.428, 0.160), which indicates that the probabilities of information flow from stakeholder B to D, E, and DI are 0.412, 0.428, and 0.160, respectively. The summation of the three probabilities equals to 1.

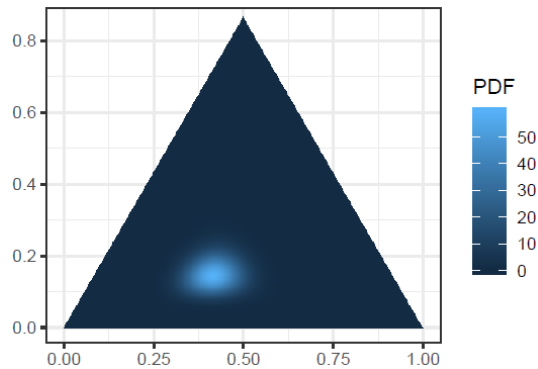

Figure 4: Probability Density Plot for $\text{Dir}(31,21,11)$.

Using equation (3) and frequencies of information flow in Figure 3, probability distributions of information flow among all stakeholders and absorbing states are computed. For one iteration, one random sample is selected from each distribution. The specialized absorbing Markov chain constructed using these probabilities is shown in Figure 5.

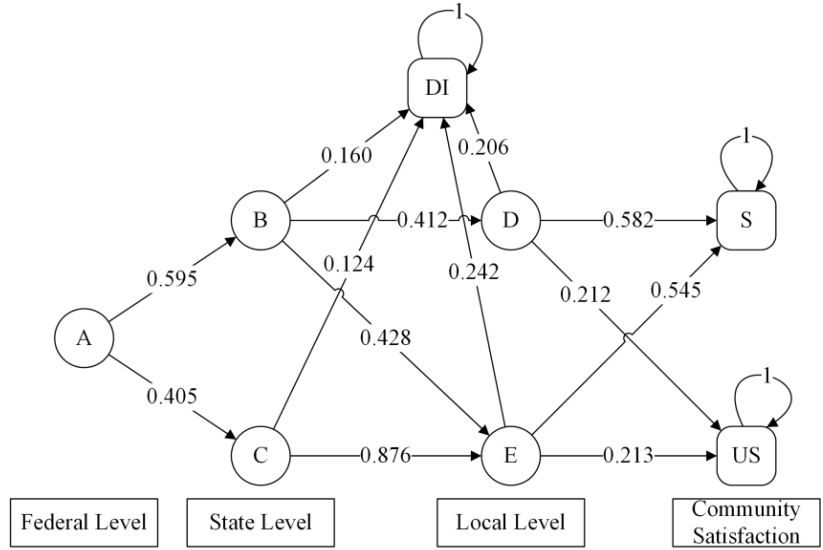

Figure 5: Absorbing Markov chain model of one iteration.

Applying equation (5), $P_{AS} = 0.475$. This value indicates that, for this iteration, the effectiveness of delivering assistance provided by federal stakeholder A to the community is 0.475. To obtain frequency histogram of $P_{AS}$, the specialized absorbing Markov chain model is run 1000 iterations. The simulated frequency histogram of $P_{AS}$ is shown in Figure 6.

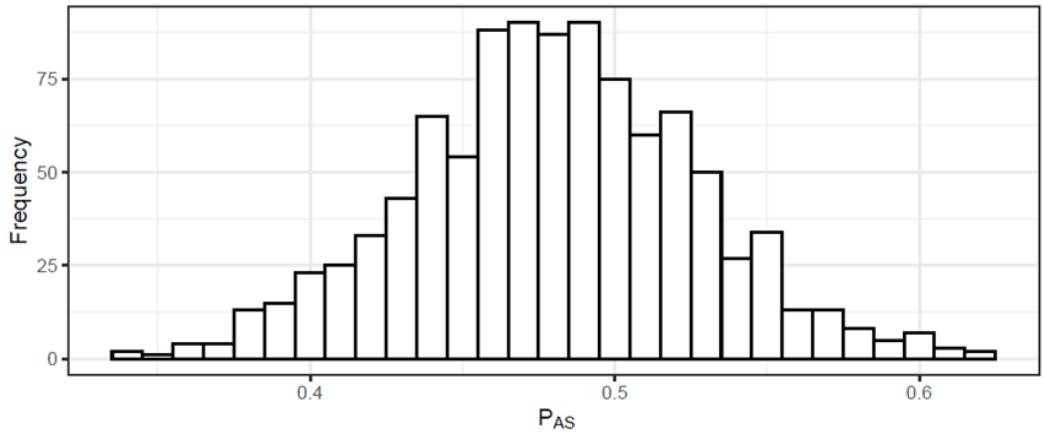

Figure 6: Frequency histogram of $P_{AS}$.

The mean value of $P_{AS}$ is 0.481, which indicates the expected effectiveness of delivering federal assistance to the community is 0.481. Based on $P_{AS}$, stakeholders can make modifications (e.g., addition or reduction of frequency of information flow) to observe changes of $P_{AS}$. Through observing $P_{AS}$ critical stakeholders can be identified, so that targeted proactive actions can be taken for improved effectiveness of delivering federal assistance to the community.

## 5    DISCUSSION

Among all types of information flow, the information flow from stakeholders to absorbing state DI causes the most impact on the effectiveness of assistance delivery because it is an indication of ineffective information flow. To verify that the specialized absorbing Markov chain model has such impact, frequencies of stakeholders' ineffective information flow (i.e., information flow from stakeholders to absorbing state DI) are adjusted to observe changes of $P_{AS}$. For demonstration purposes, stakeholder D is selected. The frequency of information flow from stakeholder B to stakeholder D is 30; therefore, the highest frequency of ineffective information flow for stakeholder D is 30. In the adjustment, frequencies of ineffective information flow for stakeholder D is increased from 0 to 30 with an increment of 1. Meanwhile, frequencies of information flow from stakeholder D to its other interacting states (i.e., absorbing states S and US) are adjusted proportionally to the original frequencies of information flow. For each increment of ineffective information flow, the specialized absorbing Markov chain model is run 1000 times and mean values of the probabilities of the process (delivering federal assistance to the community) ending up in any absorbing states $P_{AS}$, $P_{AUS}$, and $P_{ADI}$ are computed. Changes of $P_{AS}$, $P_{AUS}$, and $P_{ADI}$ resulted from each increment of ineffective information flow are illustrated in Figure 7. The effectiveness of delivering federal assistance to the community is the highest ($P_{AS} = 0.507$) when the frequency of ineffective information flow is 0, and the lowest ($P_{AS} = 0.345$) when the frequency of ineffective information flow is 30. As frequencies of information flow increase, values of $P_{ADI}$ increase, values of $P_{AS}$, and $P_{AUS}$ decrease, the summation of the three probabilities at each ineffective information flow increment equals to 1. The trend of changes in $P_{AS}$ indicates that the higher the frequency of ineffective information flow, the lower the effectiveness of delivering federal assistance to the community. The impact of ineffective information flow for stakeholder D on the effectiveness of delivering federal assistance to the community is computed as follows.

$$\frac{\Delta P_{AS}}{\Delta N_{DI}} = \frac{0.507 - 0.345}{30 - 0} = 0.00540$$

This value indicates that when stakeholder D discards information for one time, the effectiveness of delivering federal assistance to the community decreases by 0.00540. The higher the ratio is, the more impact the ineffective information flow on the effectiveness of delivering federal assistance to the community.

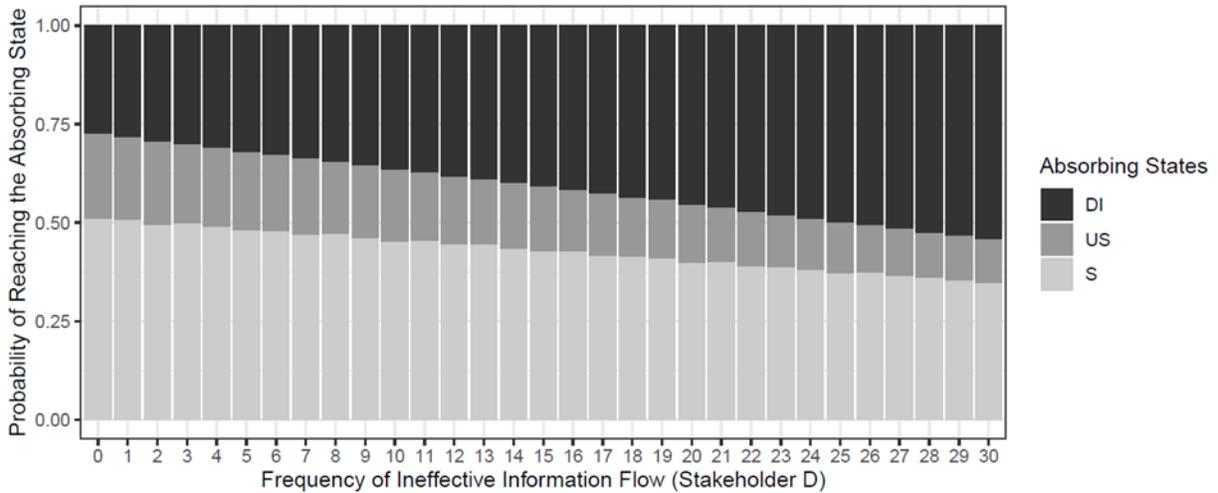

Figure 7: Impact of ineffective information flow (stakeholder D) on $P_{AS}$, $P_{AUS}$, and $P_{ADI}$.

The impact of ineffective information flow on the effectiveness of assistance delivery for stakeholder B, C, and E are computed following the same steps. The results are recorded in Table 1.

Table 1: Impact of ineffective information flow on the effectiveness of delivering federal assistance to the community.

| Stakeholder | $N_{DI(min)}$ | $N_{DI(max)}$ | $P_{AS(min)}$ | $P_{AS(max)}$ | $\Delta P_{AS}/\Delta N_{DI}$ |
|---|---|---|---|---|---|
| B | 0 | 60 | 0.535 | 0.227 | 0.00513 |
| C | 0 | 40 | 0.519 | 0.269 | 0.00625 |
| D | 0 | 30 | 0.507 | 0.345 | 0.00540 |
| E | 0 | 55 | 0.554 | 0.154 | 0.00727 |

In Table 1, $N_{DI(min)}$ and $N_{DI(max)}$ represent the minimum and the maximum frequency of ineffective information flow. $P_{AS(min)}$ and $P_{AS(max)}$ represent the effectiveness of disaster response corresponding to the minimum and the maximum frequency of ineffective information flow. Based on all these values, $\Delta P_{AS}/\Delta N_{DI}$, which reflects the impact of ineffective information flow on the effectiveness of delivering federal assistance to the community, is obtained. Using these values, stakeholders ranked from the highest impact to the lowest impact follows the order E, C, D, B. Among all stakeholders, stakeholder E has the highest value, which indicates ineffective information flow of stakeholder E has the highest impact on the responding process. Based on this result, it can be concluded that, to improve the effectiveness of the assistance delivery process, stakeholder B needs to flow less information to stakeholder E and stakeholder E needs to take proactive actions to prevent itself from discarding information.

## 6   CONCLUSION

To ensure efficient delivery of federal assistance to the community, this research derives a quantitative model to evaluate the impact of information flow on the effectiveness of disaster response. At the core of the quantitative model is a specialized absorbing Markov chain that models the process of stakeholders delivering federal assistance to the community. The model further incorporates information flow uncertainty using the Bayesian-based Dirichlet-multinomial model to enable the dynamic updating of information flow as more observations are collected, thereby measuring the effectiveness of response processes more accurately and reliably. Build upon the specialized model, the probability of community satisfaction is computed to reflect the disaster response effectiveness. In practice, governmental stakeholders can utilize the model to identify stakeholders with poor information flow performances. Using the results, targeted proactive actions can be taken for enhanced disaster response. In addition, the model has the potential to be modified for assisting stakeholders from other sectors (e.g., non-profit organizations and private businesses). The applicability and interpretability of the derived model is demonstrated using a hypothetical example. The model's performance is verified by showing a decreased effectiveness with an increased number of discarded information. Still, the model needs to be validated using practical stakeholder collaboration networks. In the future, historical disaster response information in governmental reports and agency websites will be investigated to construct the stakeholder collaboration network and to obtain information flow frequency for validating the model.

## AUTHOR BIOGRAPHIES

**YITONG LI** is a Ph.D. student in the Department of Civil, Environmental & Infrastructure Engineering, George Mason University. Yitong's current research area focuses on construction simulation modeling. Her e-mail address is yli63@gmu.edu.

**WENYING JI** is an assistant professor in the Department of Civil, Environmental & Infrastructure Engineering, George Mason University. Dr. Ji is an interdisciplinary scholar focused on the integration of advanced data analytics, complex system simulation, and construction management to enhance the overall performance of infrastructure systems. His e-mail address is wji2@gmu.edu.